\documentstyle[tighten,aps]{revtex}

\begin{document}
\draft

\title{Shear response of a frictional interface to a normal load 
modulation}

\author{L. Bureau, T. Baumberger, and C. Caroli}

\address{Groupe de Physique des Solides\cite{CNRS}, 2 place Jussieu, 
75251 Paris cedex 05, France.}

\date{\today}
\maketitle

\abstract{We study the shear response of a sliding multicontact interface 
submitted to a harmonically modulated normal load, without loss of 
contact. We measure, at low velocities ($V<100\,\mu$m.s$^{-1}$), the 
average value $\bar{F}$ of the friction force and the amplitude
 of its first and second 
harmonic components. The excitation frequency ($f=120$ Hz) is chosen 
much larger than the natural one, associated to the dynamical ageing 
of the interface. We show that: (i) In agreement with the engineering 
thumb rule, even a modest modulation induces a substantial decrease 
of $\bar{F}$. (ii) The Rice-Ruina state and rate model, though 
appropriate to describe the slow frictional dynamics, must be extended 
when dealing with our ``high'' frequency regime. Namely, the rheology 
which controls the shear strength must explicitly account not only 
for the plastic response of the adhesive junctions between 
load-bearing asperities, but also for the elastic contribution of the 
asperities bodies. This ``elasto-plastic'' friction model leads to 
predictions in excellent quantitative agreement with all our 
experimental data.
}


\section{Introduction}
\label{sec:intro}

Friction between solids carrying a time-dependent normal load is a subject 
of interest in different fields, from mechanical engineering, where the 
``friction-lowering'' effect of external vibrations is well 
known\cite{tribo} and 
commonly used in applications, to 
geophysical studies of the effect of rapid stress changes on static and dynamic 
friction of 
rocks\cite{LD,RiMa}, aiming at a better understanding of the coupling between normal 
and tangential stress states on slipping faults\cite{RiCo,RaRi,Prak}.

These studies involve 
multicontact interfaces (MCI's), {\it i.e.} interfaces between macroscopic 
solids with rough surfaces. The real area of contact thus
consists of a large number of small contacts with sizes
on the micrometer scale.\\
In a situation of constant normal load on the MCI, the 
phenomenological state- and rate-dependent friction (SRF) model, 
formulated by Rice and Ruina\cite{RR}, successfully describes the 
details of the low-velocity dynamics (typically in the 0.01 -- 100 $\mu$m.$s^{-1}$ range)
of such systems, such as the 
bifurcation between steady sate and stick-slip oscillations\cite{HeBa}. The 
model states that the 
dynamic friction force $F_{fr}$ depends on the instantaneous
sliding velocity $\dot{x}$ and on a dynamic state variable 
$\Phi$ as:
\begin{equation}
	F_{fr}\left(\dot{x},\, \Phi \right)=W\left[\mu_{0}+A \ln \left(\frac{\dot{x}}{V_{0}}\right) + B 
	\ln \left(\frac{V_{0}\Phi}{D_{0}}\right)\right],
\label{eqn:RR}
\end {equation}
where $\mu_{0}$ is the dynamic friction coefficient in steady sliding 
at the reference velocity $V_{0}$, and A and B are measured to be positive 
and of typical order $10^{-2}$ (with $B>A$).

The state variable $\Phi$ can be interpreted as the 
``age'' of the MCI, {\it i.e.} the average duration of the transient 
contact between load bearing asperities. For example, in stationary sliding at 
velocity $V$, the set of 
microcontacts is destroyed and replaced by a 
new one over a characteristic sliding length $D_{0}$, and the state variable 
thus expresses as $\Phi=D_{0}/V$.\\
More generally, the model specifies the time evolution law of $\Phi$ as:
\begin{equation}
	\dot{\Phi}=1-\frac{\dot{x}\Phi}{D_{0}}
\label{eqn:age1}
\end{equation}
In Eq. (\ref{eqn:RR}) the two corrections to 
$\mu_{0}$ have distinct physical meanings: the first term describes an 
instantaneous velocity-strengthening of the interface, while the 
second expresses strengthening of the interface with 
its ``age'', which in stationary sliding, where $\Phi=D_{0}/V$, leads 
to a velocity-weakening effect.

In the spirit of the
Bowden and Tabor analysis\cite{BT}, one can write, for a MCI, the friction 
force as\cite{PRB2}:
\begin{equation}
	F_{fr}=\sigma_{s}\left(\dot{x}\right)\Sigma_{r}\left(\Phi ,\, W\right) 
\end{equation}
where $\sigma_{s}$ defines an interfacial shear strength, 
$\Sigma_{r}$ is the real area of contact between the solids, and $W$ 
is the normal load carried by the multicontact interface. The 
 age-strengthening effect is associated with the creep growth of 
the microcontact area under normal load: 
\begin{equation}
	\Sigma_{r}\left(\Phi ,\, W \right)=\Sigma_{0}\left(W\right)\left[1+m 
	\ln \left(\frac{\Phi V_{0}}{D_{0}}\right)\right],
\label{eqn:aire}
\end{equation}
$\Sigma_{0}$ --- the real area of contact at 
$\Phi=D_{0}/V_{0}$ --- exhibits a linear dependence on $W$, as 
explained by Greenwood and Williamson's model\cite{GW} of contact between rough 
surfaces. That is, the friction force obeys the Amontons law $F_{fr}\propto W$.\\
The velocity-dependent interfacial strength of the interface is described 
as: 
\begin{equation}
	\sigma_{s}\left(\dot{x}\right) = \sigma_{s0}\left[1+\eta \ln 
	\left(\frac{\dot{x}}{V_{0}}\right) \right],
\label{eqn:sigV}
\end{equation}
This form for the interface ``rheology'', discussed in detail 
in reference\cite{PRB2}, results from the thermally activated depinning of multistable 
nanometric units localized in a layer of nanometric thickness forming a junction 
between micrometric asperities.

Eqs. (\ref{eqn:aire}) and (\ref{eqn:sigV}) yield Eq. 
(\ref{eqn:RR}) with $\mu_{0}=\sigma_{s0} \Sigma_{0}/W$, 
$m=B/\mu_{0}$, $\eta=A/\mu_{0}$, and since $m,\, \eta \ll 1$, non-linear 
logarithmic terms can be neglected.

The SRF model, and its physical interpretation  presented above, have been 
validated by friction experiments on different classes of contacting 
materials, namely granite, paper, polymer glasses and elastomers, under
constant normal load applied to the solids.

In the case of a time-dependent normal load, one can first note that 
in the Amontons-Coulomb description ($F_{fr}=\mu W$, with constant $\mu$), 
a change in $W$ would lead to a proportional change in $F_{fr}$, in particular a 
harmonic normal load modulation $W=W_{0}\left(1+\epsilon\, 
cos\left(\omega t\right)\right)$ would produce a harmonic frictional 
modulation about a non-modified average value $\mu W_{0}$.

In the SRF 
framework, the variations of 
$\dot{x}$ and $\Phi$ are non-linearly coupled, through Eqs. 
(\ref{eqn:RR}) and (\ref{eqn:age1}), to the load modulation, thus resulting in 
non-trivial 
effects on the friction force (such as, for instance, an anharmonic response to 
a harmonic normal load). However, the model as expressed by 
Eqs. (\ref{eqn:RR}) and (\ref{eqn:age1}) may not be sufficient 
to describe 
correctly the frictional response for the following reasons: 

(i) the interface rheology  expressed by Eq. 
	(\ref{eqn:sigV}) may 
	not hold for ``fast'' changes of $W$,
	
(ii) the load variation may modify the interface age strengthening 
	process, thus leading to changes in the evolution of
	 $\Sigma_{r}$ with $\Phi$, or in the evolution law of the state variable 
	 $\Phi$ itself.

Based on their results on the response to normal stress steps and pulses in granite 
friction experiments, Linker and Dieterich\cite{LD} suggested to modify the 
evolution law of $\Phi$, while retaining the functional form 
(\ref{eqn:aire}) of the $\Phi$ dependence of $\Sigma_{r}$. Argueing that a sudden change in normal 
stress would result in a sudden change in $\Phi$, they postulate:
\begin{equation}
	\dot{\Phi}=1-\frac{\dot{x}\Phi}{D_{0}}-\frac{\alpha \dot{\sigma}}{B 
	\sigma}\Phi,
\label{eqn:age2}
\end{equation}
where they infer  
$\alpha=0.2$, for granite, from their analysis of the response to 
sudden normal load steps.

In a recent study, Richardson and Marone\cite{RiMa} investigated the influence of normal stress 
modulations on the so-called ``frictional healing'' effect in a granular material 
layer confined between rough granite blocks: starting from
steady sliding, shear loading is stopped and 
the subsequent shear stress relaxation
is measured in presence of a 1 Hz  
normal load modulation (a modulation frequency close to the 
characteristic stick-slip oscillations frequency that can be inferred 
for their system).

Friction experiments with confined granular media have been successfully 
described by the SRF model in situations of constant 
normal load\cite{Marone} (although the physical meaning of the 
variable $\Phi$ is not clear yet for these systems). However, the 
use of the constitutive law proposed by Linker and Dieterich to include 
time-dependence of the normal load did not account properly for the 
details of the results obtained by Richardson and Marone.

In this paper we present an extensive study of the effect of a harmonic 
modulation of the normal load, $W=W_{0}\left(1+\epsilon \,\cos \left(\omega 
t\right)\right)$, 
on the dynamic frictional response of a multicontact 
interface. Experiments are conducted on an interface between two 
blocks of poly(methyl-metacrylate) (PMMA),
at velocities $V<100\, \mu m.s^{-1}$, load modulation frequency 
$f=120\, Hz\gg V/(2\pi D_{0})$ and relative amplitude $\epsilon$ 
in the range $5.10^{-3}\, -\, 0.5$ (so that no loss of contact 
between the surfaces occurs).

We study quantitatively the average $\bar{F}$ and the components 
at frequency $f$ and $2f$, $F_{1}$ and $F_{2}$, of the 
tangential pulling force:
\begin{equation}
$$
F=K(Vt-x)=m\ddot{x}+F_{fr}
$$ 
\label{eqn:mgamma}
\end{equation} 
for 
different values of $V$ and $\epsilon$ ; these results are presented in 
section \ref{sec:experiments}. We find in particular that the modulation 
of $W$ induces a systematic decrease of the average dynamic friction 
coefficient $\bar{\mu}=\bar{F}/W_{0}$. This effect, which increases with 
higher $\epsilon $, is quite substantial: a typical 
magnitude of this effect is a $20 \% $ decrease of $\bar{\mu}$ for 
$\epsilon=0.5$.

To analyze quantitatively our experimental data, we need to evaluate 
which fraction, $\epsilon_{eff}/ \epsilon$, of the load modulation is effectively borne by the 
microcontacts. Indeed, the normal load modulation is too fast for 
air to be drained in and out of the micrometer-thick interfacial gap.
We have studied this ``leaking air cushion'' effect by conducting similar experiments 
under primary vacuum. From these experiments we infer that $\epsilon_{eff}/ 
\epsilon \approx 0.4$, and we use this in the subsequent analysis as a 
scaling factor for 
the modulation amplitude.

Section \ref{sec:discuss} is devoted to the analysis of these results 
in terms of the SRF model and its possible extensions to fast load 
modulations:

	(i) We first test the unmodified SRF model by setting in Eq. (\ref{eqn:RR}) 
	 $W\rightarrow W(t)$ and using the evolution law (\ref{eqn:age1}) for the 
	state variable $\Phi$. Numerical integration of these equations leads 
	to a quantitatively good prediction of the average friction 
	force $\bar{F}\left(\epsilon \right)$.
	However, the predicted oscillating tangential force components $F_{1}$ 
	and $F_{2}$ strongly depart from the observed dependences on $\epsilon$ and 
	$V$.
	
	(ii) We have then tested the proposition of Linker and Dieterich.
	Using their evolution law (\ref{eqn:age2}) and their proposed value of 
	$\alpha=0.2$, we find that (i) the decrease of $\bar{F}(\epsilon)$ is much 
	smaller than the measured values and (ii) the agreement for $F_{1}$ 
	and $F_{2}$ is not better than in the previous $\alpha=0$ test.\\
	An attempt with a value of $\alpha$ small enough ($\alpha=0.02$) to describe correctly 
	the $\epsilon$ dependence of $\bar{F}$ leads to results close to those obtained 
	with the basic SRF equations; this also holds for $F_{1}$ and $F_{2}$. 
	That is, as confirmed by a perturbation 
	calculation in $\epsilon$, our experiments are not discriminating 
	with respect to the Linker-Dieterich evolution law for such small values of 
	$\alpha$.\\
	So, this modified ageing law, even if 
	valid, does 
	not suffice to account properly for the details of the frictional response.
	
	(iii) We propose to modify  expression (\ref{eqn:sigV}) 
	 for the following physical reason: we know 
	from static measurements\cite{BaBe} 
	that a MCI exhibits, at shear 
	forces much smaller than the static threshold, an elastic tangential 
	response. One can deduce from 
	this a shear stiffness $\kappa_{asp}$ with 
	the particular feature $\kappa_{asp} \propto W$. Now, in our interpretation of 
	friction, the rate variable appearing in $\sigma_{s}$ must be the 
	true
	rate of irreversible (plastic) strain of the interfacial junction of 
	nanometer thickness $h$. When taking into account the asperity 
	elasticity $\kappa_{asp}$, strictly speaking, this strain rate reads 
	$h^{-1}d\left(x-F/\kappa_{asp}\right)/dt$. In quasi stationary 
	motion, this reduces to the $\dot{x}/h$ strain rate, hence the 
	usual $\sigma_{s}(\dot{x})$ expression. In the present experimental 
	situation, $\kappa_{asp}$ is modulated as $W$ itself, and the difference 
	between the total and plastic strain rates becomes relevant. Indeed, 
	we show that this extended phenomenological elastoplastic 
	generalization of interfacial dissipation leads to a very satisfactory
	description of the average and oscillating shear responses to 
	fast normal load modulations.
	

\section{Experiments and results}
\label{sec:experiments}

\subsection{Experimental setup and methods}
\label{subsec:manip}

The tribometer is composed of a slider of mass $M$ driven along a track
through a loading spring of stiffness $K$, one end of which is
pulled at constant velocity $V$, as schematized in the inset of
Fig. \ref{fig:setup}. The slider and track are made of PMMA\cite{NBP}
with nominally
flat surfaces lapped with SiC powder to a roughness of order $1\, \mu$m,
thus forming a multicontact
interface. \\
A detailed drawing of the setup is given in Fig. \ref{fig:setup}. We
impose the
velocity $V$ of the loading point, in the range $0.1\,- \, 100\,
\mu$m.$s^{-1}$,
by means of a translation stage driven by a stepping motor.
The tangential load is applied on the slider
through a leaf spring of stiffness $K = 0.2$ N.$\mu$m$^{-1}$,
which is the more
compliant part of the system. The dead weight of the slider
is 16 N. The average normal load $W_{0}$ can be set in the range $3 -
16$ N with the help of a vertical spring attached to a
remote point itself translated horizontally at the pulling velocity
$V$ through a second translation stage, in order to prevent any
tangential coupling.

The normal load modulation is achieved by means of a vibration exciter
rigidly attached to the slider: a harmonic voltage input of given amplitude
and frequency $f$ results in a harmonic vertical motion of the moving
element of the exciter on which an accelerometer is fixed.
An acceleration  of amplitude $\gamma$
of this moving element of mass $m$ induces a normal load
modulation on the slider of amplitude $m\gamma$ at frequency $f$.
We thus obtain a normal load
$W=W_{0}\left(1+\epsilon\, \cos(\omega t)\right)$ with $\omega=2 \pi f$
and
$\epsilon=m\gamma /W_{0}$ in the range $5.10^{-3}\, - \, 0.5$.

We use the loading leaf spring as a dynamometer by measuring its
deflection $\Delta X$ by means of an eddy current displacement gauge. The
tangential force applied to the slider is thus $F=K\Delta X$. We
measure the average value of the output voltage of the gauge, and use
a lock-in amplifier to measure the amplitude of the first and second 
harmonics of this output signal with respect to 
the harmonic
excitation signal. We thus characterize the shear force through its
average value
$\bar{F}$ and
its A.C. components at frequency $f$ and $2f$, of respective amplitudes
$F_{1}$ and $F_{2}$.

The experiments are conducted according to the following protocole:
for a fixed set of parameter values $W_{0}$ and $V$ leading to steady
sliding when $\epsilon=0$, we measure
$\bar{\mu }\left(0 \right)=F\left(\epsilon=0\right)/W_{0}$.
The normal
load modulation is then set at amplitude $\epsilon$, while sliding,
and shear force measurements yield $\bar{\mu}=\bar{F}/W_{0}$,
$\mu_{1}=\left\vert F_{1} \right\vert /W_{0}$ and $\mu_{2}=\left\vert F_{2}
 \right\vert
/W_{0}$. The modulation is then
switched off and $F\left(\epsilon=0\right)$ is systematically
remeasured before setting a new value of $\epsilon$, in order to
check that no drift occurred during the measurement.
Moreover, we check that for $\epsilon \neq 0$ the shear force signal
does not exhibit low-frequency stick-slip oscillations.

The experimental results reported below have been obtained with an average
load $W_{0}=7$ N and
modulation frequencies $f$ of $120$ or $200$ Hz, chosen to be away
from any mechanical resonance frequency of the setup.


\subsection{Results}
\label{subsec:results}

\subsubsection{Average dynamic friction}

The effect of the normal load modulation on the average tangential force
response is to systematically lower the dynamic friction coefficient.
The ratio $\bar{\mu}=\bar{F}/W_{0}$ decreases as the modulation
amplitude $\epsilon$ is increased. The variation
$\Delta \bar{\mu}\left(\epsilon \right)=\bar{\mu}\left(\epsilon
\right)-\bar{\mu}(0)$, plotted on Fig. \ref{fig:mu0}, becomes larger than
the experimental noise for $\epsilon \gtrsim 0.05$, and is then quasi linear
with $\epsilon$,
though it does not extrapolate to 0 at $\epsilon=0$.

Fig. \ref{fig:mu0(V)} displays  measurements of
$\bar{\mu}\left(V\right)$  for different values of $\epsilon$. It appears
that the only effect of an increase of the load modulation amplitude is to
shift down
the $\bar{\mu}(V)$
curve, without changing the slope $\partial \bar{\mu}/ \partial
\ln(V)$.
Therefore, within experimental accuracy, $\Delta \bar{\mu}\left(\epsilon
\right)$ is
velocity-independent.

\subsubsection{A.C. components of the force response}

The oscillating force response to a load modulation
at frequency $f$ is found to be weakly anharmonic. We characterize it by
the amplitudes of the first and
second harmonics  $\mu_{1}$ and $\mu_{2}$. The ratio
$\mu_{2}/\mu_{1}$ lies typically in the range 0.1 --- 0.2.

The reduced first harmonic $\mu_{1}=\vert F_{1}\vert /W_{0}$
 increases
monotonically with $\epsilon$ and
does not show any measurable dependence on the
driving velocity, as presented on Fig. \ref{fig:mu12}.a where we plot
results at $V=1$ and $10\, \mu$m.s$^{-1}$. $\mu_{1}$ is of order
$10^{-3}$ at $\epsilon=0.5$, {\it i.e.} two orders of magnitude lower than
the average shift $\Delta \bar{\mu}$.

The amplitude of the second harmonic in the shear force response also
exhibits a monotonic increase with the modulation amplitude. Moreover,
$\mu_{2}(\epsilon)$ depends significantly on velocity, the
measured amplitude of this component being lower for smaller $V$, as
presented on Fig. \ref{fig:mu12}.b.

\subsubsection{Role of the interfacial air layer}

All the above results correspond to experiments performed at
atmospheric pressure. The PMMA surfaces in contact are nominally flat
over typically $\Sigma_{0}=7\times 7$ cm$^{2}$ but their roughness
implies
that air is trapped in an interfacial gap of
micrometric thickness $h_{0}$. Any increase in normal load is borne in 
parallel by the microcontacts and by the interfacial air layer. This 
excess pressure leads the air to leak out of the edges of the sample, 
the rate of flow being limited by the air viscosity. For instance, 
when trying to lift the slider from the track, a strong succion is 
experienced. One may therefore expect that the air layer plays a non 
negligible role in the interfacial response to load modulation.

In order to quantify experimentally this ``leaking air cushion'' effect,
we conducted a set
of control experiments under vacuum. The setup described in section
(\ref{subsec:manip}) was placed in a vacuum chamber allowing to work at
pressures down to $1$ mbar (a pressure at which the mean free path of
air molecules becomes of order $10\,\mu$m, {\it i.e.} much larger than 
the interfacial gap,
ensuring that the air effect has become negligible).

We first measure the average dynamic friction coefficient under
constant normal load and find $\mu_{0}\approx 0.5$, a value
equal to the friction coefficient at atmospheric pressure. 
This confirms that when the interfacial air layer is simply
sheared, the corresponding viscous
force is negligible with respect to the solid friction one.

Then, following the protocole described in section \ref{subsec:manip}, we
measure $\bar{\mu}\left(\epsilon \right)$, at $V=10\, \mu$m.s$^{-1}$ and
$f=200$ Hz. We have not been able to use in this control experiment the 
frequency $f=120$ Hz at which all other data have been obtained, due 
to the presence of a spurious mechanical resonance of the vacuum 
chamber close to 120 Hz.

A comparison of the average friction coefficient variation $\Delta
\bar{\mu}$ measured at $P=1$ atm and at $P=1$ mbar is
presented on Fig. \ref{fig:mu0vide}. Note that for a given
modulation amplitude, $\left\vert \Delta \bar{\mu}\right\vert$ is larger 
in vacuum than
in air. Moreover, when plotted as a function of
$\epsilon_{eff}=\epsilon/2.5$, the results obtained at $P=1$ atm
are
found to
collapse on those at $P=1$ mbar (see Fig. \ref{fig:mu0vide}).

We present in the Appendix a model calculation of the 
elastohydrodynamic response of the air layer. We show that, in all the 
range of $\epsilon$ used in our experiments, the normal response of 
the interface is linear, hence the ratio $\epsilon_{eff}/\epsilon$ 
does not depend on $\epsilon$. 
Moreover, the estimated order of magnitude of this parameter at $f=200$ Hz is 
found to be compatible with the above measured value.


\section{Discussion and model}
\label{sec:discuss}

In this section we analyse our data within the SRF framework.
The three parameters $A$, $B$ and $D_{0}$ involved in the SRF laws are
 determined experimentally, at constant load $W$, using the velocity
dependence of the friction coefficient
$\bar{\mu}\left(\ln(V)\right)$ and the dynamic characteristics of the
response close to the bifurcation threshold (this
method has been described in detail in reference \cite{PRB2}). We measure for our
system $A= 0.013\pm 0.005$, $B= 0.026\pm 0.01$ and $D_{0}= 0.4
\pm 0.04\, \mu$m. All the numerical integrations of SRF laws presented below are
performed
with this set of parameter values.

\subsection{Rice and Ruina's model}
\label{sub:RR}
Before coming to the question of whether or not the
Rice-Ruina (RR) equations themselves should be modified in the
presence of load modulations, it is reasonable to study first which
response is predicted by the RR model as such.

Replacing in Eq. (\ref{eqn:RR}) $W$ by its instantaneous value,
the equation of motion of the center of mass of the slider reads:
\begin{equation}
	M\ddot{x}=K\left(Vt-x\right)-W_{0}\left(1+\epsilon \cos(\omega
	t)\right) \left[ \mu_{0}+A \ln\left( \frac{\dot{x}}{V_{0}}\right)+
	B\ln \left(\frac{\Phi V_{0}}{D_{0}}\right) \right]
\label{eqn:move}
\end{equation}
where $x(t)$ is the instantaneous position of the center of mass of
the slider with respect to the track. We assume
the evolution law of $\Phi$ (Eq. \ref{eqn:age1}) to be unmodified:
\begin{equation}
	\dot{\Phi}=1-\frac{\dot{x}\Phi}{D_{0}}
\label{eqn:ageRR}
\end{equation}
\subsubsection{The perturbative regime}
\label{subsub:perturb}

Let us first consider the case where $\epsilon\ll 1$.
We linearize Eqs. (\ref{eqn:move}) and (\ref{eqn:ageRR})
about the steady sliding state at velocity $V$, $\epsilon=0$:
\begin{equation}
\Phi_{st}=D_{0}/V \; \; \; \;  x_{st}=Vt-W_{0}/K
\end{equation}
Setting:
\begin{equation}
	\delta \Phi= \Phi-\Phi_{st}=\Re e\left(\epsilon X_{1}\exp \left(i \omega
	t\right)\right)
\end{equation}
\begin{equation}
	\delta x=x-x_{st}=\Re e\left(\epsilon \Phi_{1}\exp \left(i \omega
	t\right)\right)
\end{equation}
we get to first order in
$\epsilon$
\begin{equation}
	\left(-M\omega^{2}+K+iW_{0}\omega \frac{A}{V}\right)X_{1} +
	W_{0}\frac{BV}{D_{0}}\Phi_{1}=-W_{0}\bar{\mu}(V)
\label{eqn:linmove}
\end{equation}
\begin{equation}
	i\frac{\omega}{V}X_{1}+\left(i\omega+\frac{V}{D_{0}}\right)\Phi_{1}=0
\label{eqn:linage}
\end{equation}
where
$\bar{\mu}\left(V\right)=\mu_{0}-\left(B-A\right)\ln\left(V/V_{0}\right)$.
We thus obtain
\begin{equation}
	X_{1}=-\frac{W_{0}\bar{\mu}}{\Delta}\left(i\omega+\frac{V}{D_{0}}\right)
\label{eqn:X1}
\end{equation}
\begin{equation}
	\Phi_{1}=\frac{i\omega W_{0}\bar{\mu}}{V\Delta}
\label{eqn:phi1}
\end{equation}
where $\Delta$ reads
\begin{equation}
	\Delta=\frac{KV}{D_{0}}\left(\frac{\omega}{\omega_{c}}\right)^{2}
	\left[-\frac{K}{K_{c}}+\left(1-\left(\frac{\omega}{\omega_{0}}\right)^{2}
	\right)
	\left(\frac{\omega_{c}}{\omega}\right)^{2}
	-i 
	\left(\frac{\omega_{c}}{\omega}\right)\sqrt{\frac{B-A}{A}}
	\left(\left(\frac{\omega}{\omega_{0}}\right)^{2}-\left(1-\frac{K}{K_{c}}
	\right)\right)\right]
\label{eqn:delta}
\end{equation}
and (see reference\cite{PRB2})
\begin{equation}
K_{c}=\frac{(B-A)W_{0}}{D_{0}}
\label{eqn:Kc}
\end{equation}

\begin{equation}
\omega_{c}=\sqrt{\frac{B-A}{A}}\frac{V}{D_{0}}
\label{eqn:omegac}
\end{equation}
are, respectively, the critical stiffness and  pulsation at the stick-slip 
bifurcation for the unmodulated system.
$\omega_{0}=\sqrt{K/M}\approx 360$s$^{-1}$ is the inertial frequency.

In our experimental conditions, with $V=10\,\mu$m.s$^{-1}$ and 
$(B-A)/A\sim 1$, $\omega_{c}\simeq 25$s$^{-1}$, so $\omega_{c}\ll \omega,\; \omega_{0}$. On the other 
hand $K/K_{c}\gtrsim 1$. Then 
$\Delta\approx-\omega^{2}AW_{0}/V$, indicating in particular that 
inertia is negligible.
Finally:
\begin{equation}
\mu_{1}=\frac{K \left\vert X_{1}\right\vert}{W_{0}}\epsilon=
\frac{\epsilon\bar{\mu}}{A}
\frac{V}{\omega}\frac{K}{W_{0}}
\label{eqn:mu1lin}
\end{equation}

Similarly, a second order expansion in $\epsilon$ yields the 
corrections at frequencies $2\omega$ and $0$, namely:
\begin{equation}
\mu_{2}\approx
\frac{V}{\omega}\frac{K\bar{\mu}^{2}}{2W_{0}A^{2}}\epsilon^{2}
\label{eqn:mu2lin}
\end{equation}
\begin{equation}
\Delta \bar{\mu}\approx
-\frac{\bar{\mu}\left(\bar{\mu}+2A\right)}{4A}\epsilon^{2}
\label{eqn:mubarlin}
\end{equation}
Note that Eq. (\ref{eqn:mubarlin}) correctly predicts a 
decrease of the average friction coefficient.

It is interesting to compare the relative perturbative corrections on 
the age and velocity variables. One finds 
\begin{equation}
\left\vert \frac{\delta \Phi/\Phi_{st}}{\delta\dot{x}/V}\right\vert \approx
\frac{\omega_{c}}{\omega}\ll 1
\end{equation}
showing that in our regime, the modulation of the age variable 
contributes negligibly to the shear response. 

Moreover, due to the smallness of $A$, the effective perturbation parameter 
is given by:  
\begin{equation}
\left\vert\frac{\delta\dot{x}}{V}\right\vert=\frac{\epsilon\bar{\mu}}{A}
\sim 50\epsilon
\label{eqn:varV}
\end{equation}
that is, the perturbative regime ($\epsilon\ll10^{-2}$) is in practice out 
of experimental reach. We thus must resort to full integration of the 
above RR equations.

\subsubsection{Average friction coefficient decrease}
\label{subsub:RRnum}

Numerical results for $\Delta
\bar{\mu}(\epsilon)$ are plotted on Fig. \ref{fig:mu0RR}
 at excitation frequency $f=120$ Hz and velocities $V=1$ and
$10\, \mu$m.s$^{-1}$. Note that a very
weak dependence on $V$ is predicted, as
observed
experimentally.

In section \ref{subsec:results} we emphasised the role played by the
interfacial air layer in our experiments, and pointed out that it 
should be taken into account
through an effective modulation amplitude $\epsilon_{eff}$, 
accounting for the fact that only a part of the excitation is borne by 
the contacting asperities. Therefore, the modulation parameter 
$\epsilon$ introduced in Eq. (\ref{eqn:move}) must be understood 
as $\epsilon_{eff}$.

On the other hand, we have measured, at $f=200$ Hz, 
$\epsilon_{eff}=0.4\,\epsilon$. As explicited in the appendix, we 
expect $\epsilon_{eff}/\epsilon$ to exhibit some relatively weak frequency 
dependence. This effect depends crucially on 
the interfacial normal stiffness which is difficult to measure 
accurately. So, we have chosen to treat $\epsilon_{eff}/\epsilon$ as a 
free fitting parameter with an initial trial value 0.4.

Fig. \ref{fig:mu0RRfit} shows the best fit obtained for  
$\Delta\bar{\mu}(\epsilon)$ at $V=1\; \mu$m.s$^{-1}$. It corresponds to 
$\epsilon_{eff}=0.48\,\epsilon$.

From now on, all experimental data will be plotted versus this 
effective modulation parameter.

\subsubsection{A.C. response}
\label{subsub:RRAC}

The computed first and second harmonics of the frictional response are
plotted on Figs. \ref{fig:muRR}.a and \ref{fig:muRR}.b.
One can first
notice that the
quasi-linear dependence of $\mu_{1}$ and $\mu_{2}$ on $V$ predicted by the
perturbation calculation also holds here. Moreover, both computed harmonics
saturate at large $\epsilon$. None of these features agrees with the 
experimental behaviour.

We therefore conclude that, in spite of the excellent agreement between the 
predicted and observed $\Delta\bar{\mu}\left(\epsilon,\, V\right)$, the 
unmodified RR model is insufficient to describe 
the full response of the interface.


\subsection{Linker and Dieterich's ageing law}
\label{sub:LD}

As mentioned in section \ref{sec:intro}, Linker and Dieterich\cite{LD} (LD) 
have proposed an extended version of the RR model in which the 
evolution law of the age variable $\Phi$ is modified according to Eq. 
(\ref{eqn:age2}).
We now study the predictions of this extended model.

A perturbation calculation to first order in $\epsilon$,
using the equation of motion (\ref{eqn:move}) and the age law
(\ref{eqn:age2}), leads to a first harmonic
amplitude:
\begin{equation}
\mu_{1}=\frac{\epsilon\vert\bar{\mu}-\alpha\vert}{A}
\frac{V}{\omega}\frac{K}{W_{0}}
\label{eqn:mu1lin2}
\end{equation}
This expression points to the fact that the dimensionless LD 
parameter $\alpha$ acts as a correction to the bare dynamic friction 
coefficient $\bar{\mu}$. LD propose for granite $\alpha=$0.2 -- 
0.3, {\it i.e.} a sizeable fraction of $\bar{\mu}$ ($\approx 0.6$ for 
that material). This leads one to expect that such a value should 
induce significant effects on the predicted shear response. However, 
we have estimated\cite{CaBa} an order of magnitude of $\alpha$ for a sparse 
population of microcontacts (Greenwood interface\cite{GW}) ageing under normal 
load. We have considered the two limits of (i) linear viscoelastic 
and (ii) fully developped plastic creep, using parameters compatible 
with the measured value of the RR parameter $B$. Both limits lead to 
the same estimate  for $\alpha$, namely one order of magnitude smaller 
than the LD value.

In view of this discrepancy, we have chosen to perform numerical integrations 
of Eqs. (\ref{eqn:move})
and (\ref{eqn:age2}) for various values of $\alpha$ in the range 
0.02 -- 0.2.

The magnitude of the load modulation effect on 
$\Delta\bar{\mu}(\epsilon)$ depends
strongly on $\alpha$, as shown on Fig. \ref{fig:muDiet}.a. Whatever 
$\alpha$, $\Delta\bar{\mu}(\epsilon)$ remains quasi-independent of 
$V$, but for $\alpha=0.2$ it is significantly smaller than the 
experimental one.

Moreover, the dependences of $\mu_{1}$ and $\mu_{2}$ on $\epsilon$ and
$V$, shown on Figs. \ref{fig:muDiet}.b and \ref{fig:muDiet}.c, as for 
the RR model, clearly disagree with the
experimental results.

$\alpha=0.02$ is found to provide a
satisfactory fit for $\Delta\bar{\mu}\left(\epsilon\right)$. However, 
this $\alpha$ value is small enough for age effects to become 
negligible, as noticed in section (\ref{subsub:perturb}). We indeed check (Figs. 
\ref{fig:muDiet}.b and \ref{fig:muDiet}.c) that the corresponding 
$\mu_{1}$ and $\mu_{2}$ are very close to those obtained from the 
unmodified RR model.

We are thus led to conclude that:\\
(i) The LD evolution law with their proposed value of $\alpha$ does 
not agree with the experimental results.\\
(ii) The $\Delta\bar{\mu}$ data permit to set an upper limit on $\alpha$ 
without, however, allowing to check the validity of the fuctional form 
of the LD model. 
Experiments at much lower frequencies (comparable to the stick-slip 
frequency $\omega_{c}$) would be needed to answer this 
question.

\subsection{Extension of the RR model}
\label{sub:RR+}

The above analysis suggests that in our  ``high 
frequency'' regime, where the response is controlled by the velocity 
modulations, it is the ``rheological'' factor $\sigma_{s}(\dot{x})$ 
which should be modified.
As mentioned in section \ref{sec:intro}, $\sigma_{s}$ 
describes the plastic dissipation occurring in a
junction of nanometer thickness between contacting asperities, and the
rate involved in $\sigma_{s}$ is a rate of {\it irreversible} strain of
this junction.

It has been shown\cite{BaBe} that when a multicontact interface is submitted to 
a shear much smaller than the static threshold, its response is 
elastic. Since the asperity ``bodies'' (which deform on a micrometric 
thickness, of the order of their diameter) are much more compliant 
than the nanometer-thick elastically pinned adhesive joint\cite{PRB2}, 
it is their response which controls the 
interfacial shear stiffness $\kappa_{asp}$. This obeys an extented 
Amontons law: 
$\kappa_{asp}=W/\lambda$, with $\lambda$ a length of order $1\, \mu$m
for our surfaces.

Sliding amounts to depinning of the adhesive joint which becomes 
dissipative, while the bodies of the asperities retain their elasticity.
Therefore, we can schematically represent the sliding interface as an elastic
element of stiffness $\kappa_{asp}$, accounting for the bulk elastic strain
of the asperities, coupled in series to a (frictional) dissipative element 
(see Fig. \ref{fig:interface}). When this latter is sheared at 
velocity $\dot{x}_{pl}$, the corresponding force is 
$F=f\left(\dot{x}_{pl}\right)$. 
\begin{equation}
F=\kappa_{asp}\, x_{el}=f (\dot{x}_{pl})
\end{equation}
with $x_{el}$ and $x_{pl}$ respectively the elastic and irreversible
displacements.
The instantaneous velocity of the center of mass of the slider thus 
reads:
\begin{equation}
\dot{x}=\dot{x}_{el}+\dot{x}_{pl}=\frac{d\left(F/\kappa_{asp}\right)}{dt}
+f^{-1}\left(F\right)
\end{equation}
and the tangential force on the slider finally reads:
\begin{equation}
F=f\left(\frac{d}{dt}\left(x-F/\kappa_{asp}\right)\right)
\end{equation}

We therefore express the external force using the same
functional form as in Eq. (\ref{eqn:RR}), but the argument of the
 rate-dependent
term becomes $\dot{x}_{pl}$.
In stationary sliding under constant normal load, 
$F/\kappa_{asp}$ is constant, hence the usual dependence on $\dot{x}$.
In the presence of a load modulation, both $F$ and 
$\kappa_{asp}=W_{0}\left(1+\epsilon
cos\left(\omega t\right)\right)/\lambda$ are modulated, 
and the elastic strain term becomes significant.

We present hereafter the results obtained from numerical integration
of the corresponding extended RR equations. Taking into 
account the above mentioned fact that in our experimental conditions, 
inertia can be neglected, Eq. (\ref{eqn:RR}) becomes:
\begin{equation}
	F/W=\frac{K}{W}(Vt-x)=
	\mu_{0}+A \ln 
	\left[\frac{1}{V_{0}}\frac{d}{dt}\left(x-\frac{K}{\kappa_{asp}}(Vt-x)\right)
	\right] + B 
	\ln \left(\frac{V_{0}\Phi}{D_{0}}\right),
\label{eqn:RRext}
\end {equation}

The parameters $A$, $B$ and $D_{0}$ are set to their
experimentally determined values. The length $\lambda$ has been 
obtained from a quasi-static loading-unloading test\cite{BaBe} at 
various normal loads. We find $\lambda=0.62\pm 0.15\, \mu$m. In view 
of the relatively large experimental uncertainty on this parameter, 
we have integrated Eqs. (\ref{eqn:RRext}) and (\ref{eqn:age1}) with 
$\lambda$ as a free parameter. The best fit, performed on the most 
sensitive data, namely the $\mu_{1}(\epsilon_{eff})$ ones, is found to 
correspond to $\lambda=0.7\, \mu$m, within the experimental 
uncertainty braket.

While $\Delta\bar{\mu}\left(\epsilon\right)$ is found to be only very weakly 
affected by the rheological correction, this extension of the model yields 
predictions 
for $\mu_{1}$ and $\mu_{2}$ markedly different from those of both the 
unmodified RR and LD models. Namely, their quasi-linear 
$V$-dependences are replaced by much weaker ones. On the other hand, 
neither $\mu_{1}$ nor $\mu_{2}$ exhibit any longer saturation within the 
relevant range $\epsilon_{eff}<0.3$.

As appears from Figs. (\ref{fig:murheo}), 
the global agreement is now 
excellent, confirming the validity of the extended RR model.

\subsection{Concluding remarks}

This study leads us to the following conclusions:

On the one hand, from an engineering point of view, the most 
spectacular effect of modulating the normal load applied to a 
frictional system is to lower significantly the dynamic friction 
coefficient. This occurs as soon as the modulation is applied, even 
though its amplitude is low enough to ensure permanent interfacial 
contact between the sliding bodies.

On the other hand, an important aim of this study was to elucidate 
the question, relevant to seismology, of
whether the RR model should be modified to describe the frictional response to 
fast variations of the normal stress. We have shown that, in order to 
study this, it is essential to measure and analyze not only the zero 
frequency component of the response to an oscillatory load, but also 
its harmonic content.

In the range of frequencies, much larger than the stick-slip one, 
that we have studied, the shear response is controlled by the velocity 
modulation, that is by the {\it rate-dependent} term of the RR constitutive 
law. However, the quantitative analysis of $\mu_{1}$ and $\mu_{2}$ data shows 
that the relevant displacement rate {\em is not}, for fast 
load modulations, {\em the slider velocity}, but the rate of plastic 
deformation of the adhesive frictional joint.

This confirms our picture\cite{PRB2} of sliding friction as 2D plasticity
prelocalized within a nanometer-thick adhesive joint coupled to the 
bulk of the slider through elastic asperities.\\
This enables us to extend correspondingly the expression of the rate-dependent 
part of the RR state- and rate-dependent model.

The question of the precise effect of a load modulation on interfacial 
age remains at this stage open. Indeed, we have concluded 
that, at least for our system,  
this effect is certainly much smaller than proposed by Linker and 
Dieterich. However, precisely for this reason, the ``high frequency'' 
response is {\it not} a good tool for investigating this question. 
This should be adressed through similar experiments at low 
frequencies, 
close to the stick-slip one.


\appendix

\section{Elasto-hydrodynamic response of the interfacial air layer}

The aim of this appendix is to establish the equation for the vertical
motion of the slider (i.e. along the $z$-direction normal to the interface)
and to estimate the relative contributions of the forces that are involved.
We will, as a result, justify the use of an {\it effective} amplitude of
modulation of the normal load, to account for the fraction
of the modulation which is borne by the air cushion trapped within
the interfacial gap.
The order of magnitude of this fraction, referred to as
$\epsilon_{eff}/\epsilon$ in the text, and which is the only
fitting parameter of our model, is checked independently
in a control experiment, performed in vacuo, and described
in the text.

The motion of the slider along the $z$-axis is assumed to be
decoupled from the sliding motion along $x$. It is
parametrized by the width $h$ of the ``gap'', i.e. the separation
between the average planes passing through the rough surfaces of,
respectively, the track and the slider. When no modulation is
superimposed to the bare normal load $W_{0}$, the width is $h_{0}$, a
value fixed by the deformation of the load bearing asperities which
are randomly distributed over the interface of nominal macroscopic area
$\Sigma_{0}$.

\paragraph{Elastic response of the multicontact interface.}

According to Greenwood and Williamson's model for
multicontact interfaces,
the number of load bearing asperities and the real area of contact
increase linearly with the load. This induces
a non-linear dependence on load of the gap thickness.
Experimentally, it has beeen found that $h_{0}-h \simeq \lambda_{z}
\ln(W/W_{0})$,
with $\lambda_{z}$ a length, the order of magnitude of which is given by
the roughness of the surfaces in contact (the standard deviation of
the surface heights, here $1.3 \,\mu$m).
In the small amplitude linear regime ($\Delta h \ll \lambda_{z}$), the
stiffness $\kappa_{z} = W/\lambda_{z}$ is a constant and the emlastic
restoring force in the $z$-direction reads:
\begin{equation}
\label{eq:elastic}
F_{el} \simeq W_{0}\frac{(h - h_{0})}{\lambda_{z}}
\end{equation}
This expression has its exact counterpart for tangential motion, as
discussed in the text. Shear elasticity involves a length $\lambda$ which
is expected to
be about $1.7\,\lambda_{z}$ for a Poisson ration $\nu = 0.44$.
Therefore, the measured value $\lambda = 0.7
\,\mu$m yields $\lambda_{z} \simeq 0.4 \,\mu$m.

\paragraph{Elasto-hydrodynamic response of the interfacial air layer. }

When the gap is e.g. narrowed, air is compressed until being
drained out of the interfacial zone.
The resulting pressure force on the slider will be denoted $F_{p}$.
For the sake of evaluating $F_{p}$, we will simplify the problem and
consider a
thin layer of air, of viscosity $\eta$ and density $\rho_{0}$ at
atmospheric pressure $P_{0}$, trapped between two {\it perfectly
smooth}
discs of radius $R$, parallel and distant of $h \ll R$. The relative velocity
is supposed to vanish at $z=0$ and $z=h$, an assumption which is
legitimate if the roughness of the surfaces is much smaller than
the gap width.
Brown and Scholz have reported measurement of the gap
width between macroscopic ground glass surfaces. At low average pressure
corresponding to $10^{-5}$ of the Young modulus of the glass, as
encountered in our experiments, they have found that the gap width
is typically 5 times larger than the rms roughness of the statistically
identical
surfaces. This figure is clearly  too small for the
``smoothed'' model of the interfacial gap
 to be expected to provide a very accurate value of the
hydrodynamic force, though it is certainly sufficient to estimate its
order of magnitude.

An upper bound for the
average pressure excess resulting from the motion of the disk is
$\Delta P = \epsilon W_{0}/\Sigma_{0}$, with $\epsilon W_{0}$ the
amplitude of the normal load modulation and $\Sigma_{0} = \pi R^{2}$.
The macroscopic loading
pressure $W_{0}/\Sigma_{0}$ remains of order 10 mbar in the
reported experiments, while $\epsilon$ is smaller than unity. As a
result, $\Delta P$ remains much smaller than the atmospheric
pressure $P_{0}$.
However, the {\em compressibility} of the air layer may be of paramount
importance,
as suggested by the following argument. For infinite plates, no leak
occurs at the edge of the gap and the response of the layer, trapped
under the mean pressure $P_{0}$, is
elastic with an overall stiffness:
\begin{equation}
\label{eq:stiffair}
\kappa_{air} = P_{0} \Sigma_{0}/h_{0}
\end{equation}
For $P_{0} = 10^{5} Pa$, $\Sigma_{0} = 49$ cm$^{2}$ and $h_{0} =
6.5\,\mu$m, one finds $\kappa_{air} = 7.5\, 10^{7}$ N/m, namely one
order of magnitude larger than the interfacial stiffness $\kappa_{z}$
originating from the load bearing asperities (see
Eq.\ref{eq:elastic}) at $W_{0}=7$ N. For finite radius $R$,
edge flow will reduce the amount of air to be compressed in order to
accomodate the change of gap volume.
It is therefore necessary to
compute the expression of $F_{p}$ by taking account  the radial,
viscosity controlled, Poiseuille
flow which results from the density (hence pressure) gradient compatible
with mass conservation.

The continuity equation for the radial flow
reads:
\begin{equation}
\label{eq:cont}
\frac{\partial}{\partial t} (\rho h) +
\frac{1}{r}\frac{\partial}{\partial r} (r\bar{v}\rho h) = 0
\end{equation}
where $\bar{v}(r)$ is the mean velocity at radius $r$ (averaged across the gap
along the z-direction).
The pressure field is given by the equation of state of the air at
pressures close to $P_{0}$ = 1 atm, which is assumed to be:
\begin{equation}
\label{eq:state}
\frac{P}{\rho} = \frac{P_{0}}{\rho_{0}}
\end{equation}
The set of equations is closed by assuming that the flow is of the
Poiseuille type, namely is parabolic along the $z$-direction and varies
slowly along the radial direction according to:
\begin{equation}
\label{eq:poiseuille}
\bar{v} = -\frac{h^{2}}{12\eta}\frac{\partial P}{\partial r}
\end{equation}
As mentioned, the pressure modulation remains much smaller than
$P_{0}$, and the gap modulation is smaller than $h_{0}$, hence
linearization of Eqs. (\ref{eq:cont}--\ref{eq:poiseuille})
is
legitimate. One therefore sets $P = P_{0} +\delta P$, $\rho =
\rho_{0} + \delta \rho$, and $h = h_{0} +\delta h$,
with $\delta P \ll P_{0}$, $\delta\rho
\ll\rho_{0}$ and $\delta h \ll h_{0}$.
Eliminating $\delta\rho$ yields the following equation for the pressure field:
\begin{equation}
\label{eq:pressure}
\frac{h^{2}}{12\eta}\,\frac{1}{r}\frac{\partial}{\partial r}
\left(r\frac{\partial (\delta P)}{\partial r}\right)
-\frac{\dot{\delta P}}{P_{0}} =  \frac{\dot{\delta h}}{h_{0}}
\end{equation}
where the dot indicates the partial derivative with respect to time.

Assuming that the normal elastic stiffness $\kappa_{z}$ of the
asperities remains linear, the gap modulation resulting from the
normal load one is harmonic and we therefore seek for a complex solution
to Eq.\ref{eq:pressure} of the form: $\delta P = \tilde{\delta P} \exp
(i\omega t)$ with $\delta h = \tilde{\delta h} \exp
(i\omega t)$.
Taking into account the boundary condition $\delta P = 0$ at $r = R$ and the
symetry requirement $\bar{v}= 0$, hence
$\partial P/\partial r = 0$, at $r=0$, one obtains:
\begin{equation}
\label{eq:dP}
\tilde{\delta P} = -P_{0}\frac{\tilde{\delta
h}}{h_{0}}\left[1-\frac{J_{0}(\gamma r)}{J_{0}(\gamma R)}\right]
\end{equation}
With $J_{0}$ the Bessel function of zeroth order and $\gamma$ a
complex constant given by:
\begin{equation}
\label{eq:defgamma}
\gamma = \frac{1-i}{\sqrt{2}}
\sqrt{\frac{12\eta\omega}{P_{0}h_{0}^{2}}}
\end{equation}
Integration of the pressure field over the interface yields the
complex amplitude of the pressure force:
\begin{equation}
\label{eq:force}
\tilde{\delta F_{p}} = P_{0} \pi R^{2} \frac{\tilde{\delta
h}}{h_{0}} \frac{J_{2}(\gamma R)}{J_{0}(\gamma R)}
\end{equation}
with $J_{2}$ the Bessel function of second order.

The asymptotic limits deserve comments. For $\gamma R \to \infty$,
$J_{2}/J_{0}\to -1$, $\tilde{\delta F_{p}}$ is real and
one recovers the purely elastic response (Eq.\ref{eq:stiffair})
predicted for large $R$. It also corresponds to the high frequency limit
for which the air has no time to leak.
For $\gamma R \to 0$, $J_{2}/J_{0}\to (\gamma R)^{2}/8 = - i
\omega (3\eta R^{2})/(2P_{0}h_{0}^{2})$, hence  $\tilde{\delta
F_{p}}$ is purely imaginary and reduces to a linear viscous damping
force which could have been derivated by assuming a non compressive
Poiseuille flow.
For intermediate values of $\gamma R$, $\tilde{\delta F_{p}}$ has both
a reactive component, which increases the interfacial stiffness, and a
dissipative
one.

\paragraph{Prediction for the effective amplitude of load modulation.}

The slider of mass $M$ oscillates in the normal $z$-direction under the
combined
action of the load modulation, the restoring elastic force resulting
from  deformation of the load bearing asperities and compression
of the air cushion, and the damping force resulting from the air flow.
The complex amplitude of modulation of the gap width
$\tilde{\delta h}$ is therefore given by:
\begin{equation}
\label{eq:dynamique}
(-M\omega^{2} + \kappa_{z}) \tilde{\delta h} -
\tilde{\delta F_{p}} = \epsilon W_{0}
\end{equation}
with $\tilde{\delta F_{p}}$ given by Eq.\ref{eq:force}.

The fraction of the load which is effectively borne by the
microcontacts is
$\epsilon_{eff}/\epsilon = \vert\kappa_{z}\tilde{\delta h}/(\epsilon
W_{0})\vert$.
It reads:
\begin{equation}
\label{eq:epseff}
\frac{\epsilon_{eff}}{\epsilon} =
\left\vert
1 - \frac{\omega^{2}}{\omega_{0}^{2}} - \frac{\pi
R^{2}P_{0}}{W_{0}}\,\frac{\lambda_{z}}{h_{0}}\,
\frac{J_{2}(\gamma R)}{J_{0}(\gamma R)}\right\vert^{-1}
\end{equation}
with $\omega_{0} = \sqrt{\kappa_{z}/M}$.

The assumption that $\epsilon_{eff}/\epsilon$ does not depend on the amplitude
of the modulation relies upon the fact that both the elasticity and
the viscosity remain linear, namely, as previously discussed, that
$\Delta h \ll \lambda_{z}$ and $h_{0}$. This reduces to
$\epsilon_{eff}\ll 1$, a criterion which is always fulfilled in our
experiments.

For $W_{0} = 7$ N, $M = 1.6$ kg and $\lambda_{z} \simeq 0.4\,\mu$m,
$\omega_{0}/(2\pi) = 530$ Hz. Hence, at 120 Hz, the inertia
is  $5.2\,10^{-2}$ of the elastic restoring force due to the
asperities solely.
It is clear from the above analysis that a key parameter for
evaluating the viscoelastic response of the air is the gap width $h_{0}$.
Taking,  as discussed previously, the conservative value of 5 times
the roughness, namely $h_{0} = 6.5\,\mu$m, $\eta = 10^{-5}$ Pa.s, $R =
3.9$ cm, one computes $\gamma R = 4(1-i)$ and
$\vert\epsilon_{eff}/\epsilon\vert \simeq  0.24$, a value of the same
order of magnitude than the one, namely 0.48, which is found to provide
the best agreement between the experimental data and the model
prediction for $\Delta\bar{\mu}$.
The role of the interfacial air cushion is further confirmed by the
control experiment performed in vacuo. At a remaining pressure of 1
mbar, the elastic stiffness of the air layer falls two orders of
magnitude below the multicontact one. Moreover, since the mean free
path (at 300 K) of the gas molecules is now of order several 10 $\mu$m,
i.e. larger than the gap width,
the viscosity of the layer should vanish. Consequently, the effective
amplitude is essentially ruled by the slider inertia according to:
$\epsilon^{vacuum}_{eff}/\epsilon \simeq\vert 1
-\omega^{2}/\omega_{0}^{2}\vert^{-1}$ = 1.2 at 200 Hz. At atmospheric
pressure,
keeping the nominal value $\Sigma_{0} = 49$ cm$^{2}$,
$\epsilon^{air}_{eff}/\epsilon = 0.23$. When bringing the data for
$\Delta\bar\mu (\epsilon)$
performed in the
air and in vacuo  to collapse on a single curve, as explained in the
text, one makes use of a scaling ratio which reads explicitely:
$\epsilon^{air}_{eff}/\epsilon^{vacuum}_{eff} \simeq 0.20$. The
experimental value is 0.4.

The fact that, in both cases, the estimated value is
smaller than the observed one by the same amount may be
possibly
attributed to some long wavelength modulations of the gap width $h_{0}$
which is likely to remain after the lapping process. Microcontacts may
be distributed on patches of macroscopic area smaller than
$\Sigma_{0}$, separated by regions of much wider gap in which the air
would play a negligible role.
Typically, a patch radius of 2.5 cm, while keeping the other
parameters unchanged, would account for the observed value
$\epsilon_{eff}/\epsilon \simeq 0.48$ at 120 Hz in the air.
This would correspond to an effective area of $0.4\,\Sigma_{0}$, a
value still large enough for the microcontacts --- the number of which
does not depend on $\Sigma_{0}$, according to Greenwood ---
to remain elastically independent.
In addition, we have assumed a single degree of freedom for the
slider, which is certainly a strong requirement since the slider is
left free to find its own seat on the track.
It is clear that a small amount of "rocking" would promote the air
flow and reduce the cushion effect.

Finaly, normal load modulation induces a tangential oscillating
motion of the slider of
amplitude $\Delta x$,
hence an air shear flow within the gap. The associated A.C. viscous force
on the slider $\eta \omega \Delta x \Sigma_{0}/h_{0}$, which has been
neglected in our models, must be compared to
the leading term in the rate dependent friction force
for oscillations about the sliding velocity $V$,
namely $A W_{0} \omega \Delta x /V$. The ratio of both terms is $\eta
\Sigma_{0} V/(W_{0}h_{0}) \simeq 10^{-7}$ for $V = 100 \,\mu$m/s,
therefore shear viscosity of the layer is totally negligible.



\begin{figure}
\caption{Main elements of the experimental setup: 
Translation stage (Drv); Loading leaf spring (Lsp); Displacement 
gauge (Gg); Vibration exciter (Vb); Weighting spring (Spr);  
Accelerometer (Acc). The inset is the schematic representation of the 
spring-slider-track 
dynamical system with control parameters W (normal load), V (driving 
velocity) and K (external spring stiffness).}
\label{fig:setup}
\end{figure}
\begin{figure}
\caption{Variation of the reduced average friction force 
$\Delta \bar{\mu}=(\bar{F}\left(\epsilon 
\right)-F_{0})/W_{0}$ versus 
modulation amplitude $\epsilon$ at $f=120\, Hz$. 
Open symbols correspond to two sets of results at $V= 1\, \mu$m.s$^{-1}$ 
and full symbols  to two sets at $V= 10\, \mu$m.s$^{-1}$}
\label{fig:mu0}
\end{figure}
\begin{figure}
\caption{Reduced average friction force $\bar{\mu}$ $vs.$ $V$ for 
various values of 
load modulation amplitude. 
Open circles: $\epsilon=0$; Open triangles: $\epsilon=0.2$; 
Full circles: $\epsilon=0.35$; Full triangles: $\epsilon=0.5$.}
\label{fig:mu0(V)}
\end{figure}
\begin{figure}
\caption{Amplitude of the harmonic components of the reduced force 
response at $f=120$ Hz: (a) first harmonic 
$\mu_{1}(\epsilon)$; (b) second harmonic $\mu_{2}(\epsilon)$. Two sets 
of data are ploted for each velocity:  
$V=1\,\mu$m.s$^{-1}$ (open symbols) and
$V=10\, \mu$m.s$^{-1}$ (full symbols).}
\label{fig:mu12}
\end{figure}
\begin{figure}
\caption{Reduced average 
friction force $\Delta \bar{\mu} (\epsilon)$, for 
$V=10\, \mu$m.s$^{-1}$ and $f=200\, Hz$ at pressures $P$ = 1 mbar (open 
circles) and $P$ = 1 atm (open triangles). The same set of data at $P$ = 
1 atm is also plotted (full triangles) as a function of the scaled amplitude $\epsilon/2.5$.}
\label{fig:mu0vide}
\end{figure}
\begin{figure}
\caption{Predictions of the RR model for the average friction 
decrease $\Delta\bar{\mu}(\epsilon)$, 
at $f=120\, Hz$. Lines: RR model for
$V=1\, \mu$m.s$^{-1}$ (full), and $V=10\, \mu$m.s$^{-1}$ (dashed).
Symbols: raw 
experimental results at $V=1\, \mu$m.s$^{-1}$ (circles) and  $V=10\, 
\mu$m.s$^{-1}$ (triangles). 
The experimental data have been averaged over three different runs. The error bars 
correpond to standard deviations on these runs.}
\label{fig:mu0RR}
\end{figure}
\begin{figure}
\caption{Determination of $\epsilon_{eff}$ from $\Delta \bar{\mu}$. 
Full circles: raw experimental data at $V=1\, \mu$m.s$^{-1}$ and $f=120$ 
Hz.  Open circles: the same set of data plotted as a function of 
$\epsilon_{eff}=\epsilon/2.1$ which provides the best agreement with 
the RR model prediction (line).}
\label{fig:mu0RRfit}
\end{figure}
\begin{figure}
\caption{Comparison between experimental data (symbols) and	the RR model 
predictions (lines) at $f = 120$	Hz for
(a)	$\mu_{1}(\epsilon_{eff})$ and (b)	$\mu_{2}(\epsilon_{eff})$, at
$V=1\,\mu$m.s$^{-1}$ (full line and triangles), and $V=10\,\mu$m.s$^{-1}$ (dashed line and circles).}
\label{fig:muRR}
\end{figure}
\begin{figure}
\caption{Comparison between	experimental data at $f=120$ Hz	and	the LD model predictions  for	
(a) $\Delta\bar{\mu}(\epsilon_{eff})$; (b)	
$\mu_{1}(\epsilon_{eff})$; (c) $\mu_{2}(\epsilon_{eff})$. Lines:	
LD model	for
$V=1\, \mu$m.s$^{-1}$ and $\alpha=0.2$ (dotted);	$V=10\,	
\mu$m.s$^{-1}$, $\alpha=0.2$	(dashed); $V=10\, \mu$m.s$^{-1}$, 
$\alpha=0.02$ (full). The predictions for $V=1\,	\mu$m.s$^{-1}$ and 
$\alpha=0.02$ are not plotted here because they would be undistinguishable 
from	the	dotted lines.
Symbols : experimental data at 
$V=1\mu$m.s$^{-1}$ (triangles), and $V=10\mu$m.s$^{-1}$ (circles).}
\label{fig:muDiet}
\end{figure}
\begin{figure}
\caption{Schematic rheological representation of the 
frictional interface (see text).}
\label{fig:interface}
\end{figure}
\begin{figure}
\caption{Comparison between experimental data at	$f=120$	Hz (symbols) and 
the {\it	extended} RR 
model predictions (lines) for 
(a) $\Delta\bar{\mu}(\epsilon_{eff})$; (b)	$\mu_{1}(\epsilon_{eff})$; 
(c) $\mu_{2}(\epsilon_{eff})$ at	
$V=1\,\mu$m.s$^{-1}$	(full lines	and	triangles) and $V=10\,\mu$m.s$^{-1}$ (dashed lines and circles).}
\label{fig:murheo}
\end{figure}
\end{document}